\documentclass[12pt]{iopart}

\usepackage{graphicx}
\usepackage{setstack}
\usepackage{iopams}
\usepackage{color}
\usepackage[british]{babel}

\begin{document}

\title[Ergodicity breaking and ageing in generalised diffusion processes]{Ergodicity
breaking, ageing, and confinement in generalised diffusion processes with position
and time dependent diffusivity}

\author{Andrey G. Cherstvy$^\dagger$ and Ralf Metzler$^{\dagger,\sharp}$}
\address{$\dagger$ Institute for Physics \& Astronomy, University of Potsdam,
D-14476 Potsdam, Germany\\
$\sharp$ Physics Department, Tampere University of Technology, FI-33101
Tampere, Finland}

\begin{abstract}
We study generalised anomalous diffusion processes whose diffusion coefficient
$D(x,t)\sim
D_0|x|^{\alpha}t^{\beta}$ depends on both the position $x$ of the test particle
and the process time $t$. This process thus combines the features of scaled Brownian
motion and heterogeneous diffusion parent processes. We compute the ensemble and
time averaged mean squared displacements of this generalised diffusion process. The
scaling exponent of the ensemble averaged mean squared displacement is shown to
be the product of the critical exponents of the parent processes, and describes both
subdiffusive and superdiffusive systems. We quantify the amplitude fluctuations of
the time averaged mean squared displacement as function of the length of the time
series and the lag time. In particular, we observe a weak ergodicity breaking of
this generalised diffusion process: even in the long time limit the ensemble and
time averaged mean squared displacements are strictly disparate. When we start to
observe this process some time after its initiation we observe distinct features
of ageing. We derive a universal ageing factor for the time averaged mean squared
displacement containing all information on the ageing time and the measurement time.
External confinement is shown to alter the magnitudes and statistics of the ensemble
and time averaged mean squared displacements.
\end{abstract}


\section{Introduction}

Deviations from Brownian motion are quite ubiquitous in a large variety of complex
systems. Mostly such anomalous diffusion processes are characterised by a scaling
form of the mean squared displacement (MSD),
\begin{equation}
\label{msd}
\left<x^2(t)\right>\sim t^{\kappa}.
\end{equation}
The magnitude of the anomalous scaling exponent $\kappa$ distinguishes subdiffusion
for $0<\kappa<1$ and superdiffusion $\kappa>1$ \cite{bouchaud,report,havlin,pccp}.
Examples for anomalous include the relative diffusion of tracers particles in 
fully turbulent \cite{richardson} or weakly chaotic \cite{swinney} systems as
well as in groundwater aquifers \cite{brian}. Subdiffusion of charge carriers in
amorphous semiconductors was originally analysed some 40 years ago \cite{scher} but
is now regaining attention in the study of polymeric semiconductors \cite{schubert}.
However, the main current impetus for the study of anomalous diffusion processes is
due to modern spectroscopic tools such as fluorescence correlation spectroscopy
\cite{fcs} and single particle tracking of submicron particles \cite{brauchle}.
By these methods pronounced deviations from normal diffusion ($\kappa=1$) were found
in structured and crowded liquids \cite{wong,mattsson,weiss,lene1} and in living
biological cells \cite{lene,weigel,franosch}. Similarly, anomalous diffusion is
observed in supercomputer studies of, inter alia, flexible networks \cite{structure}
and lipid membranes \cite{membranes}.

The non-Brownian scaling of the MSD of the tracer particles may originate from a
range of physical mechanisms. These include the subdiffusive motion on geometric
fractals such as percolation clusters close to criticality \cite{havlin,franosch,
fractals} or the anomalous diffusion in Lorentz gases \cite{franosch_lg}. Another
important class of anomalous diffusion models are continuous time random walks
(CTRWs), in which the test particle's motion is interrupted by random waiting
times \cite{montroll}. If the distribution of these waiting times is scale free,
subdiffusion emerges \cite{scher}, a phenomenon closely related to quenched trap
models with exponentially distributed trap depths \cite{bouchaud,trap}. The third
class represents processes, in which the particle is driven by fractional Gaussian
noise which is long range temporally correlated: the fractional Brownian motion
\cite{mandelbrot} or associated fractional Langevin equation motion \cite{goychuk}
are connected to viscoelastic environments and represent the motion of tagged
monomers in a Rouse chain \cite{rouse} or a tracer particle in a single file
\cite{singlefile}. A more complete overview of anomalous diffusion models 
provides Ref.~\cite{pccp}.

Here we deal with the remaining class of popular stochastic models, in which the
diffusion anomaly stems from the explicit position or time dependence of the
diffusion coefficient. Modelling anomalous diffusion via a coordinate dependent
diffusivity goes back to Richardson's analysis of relative diffusion in turbulent
flows with $\kappa=3$ \cite{richardson}. Systematic variations of the local
diffusion coefficient in space were, for instance, demonstrated in the cytoplasm
of living biological cells \cite{lang}. A series of publications recently explored
the stochastic properties of such heterogeneous diffusion processes (HDPs)
\cite{fulinski,hdp,hdp1,hdp2,hdp3,massignan}. Instead of the position dependence,
Batchelor introduced a time dependent diffusion coefficient for the description
of Richardson diffusion \cite{batchelor}. Scaled Brownian motion (SBM) with a
power law form for the diffusivity is very popular in the phenomenological
description of anomalous diffusion \cite{saxton1}. Its stochastic properties were
analysed in detail in Refs.~\cite{fulinski,lim,thiel,sbm,hadiseh,anna}. SBM was
shown to grasp essential features of granular gases in the homogeneous cooling
phase \cite{anna1}.

Physically, HDPs and SBM are quite different: HDPs are truly Markovian processes
(with multiplicative noise \cite{fulinski,hdp}) while the time dependence of the
diffusion coefficient in SBM already heralds its fundamentally non-stationary
character. Indeed experimentally it is often not possible to separate the effects
of time and position variation of the diffusivity---compare, for instance, the
data in Ref.~\cite{platani}. This poses the question how exactly the position and
time dependence of the diffusion coefficient built into HDPs and SBM conspire. How
do they compete with each other? On the level of the stochastic Langevin we here
analyse in detail generalised diffusion processes (GDPs) with the position-time
dependent diffusion coefficient
\begin{equation}
\label{eq-diff-coeff-of-x-t}
D(x,t)\sim(1+\beta)D_0|x|^{\alpha}t^{\beta},
\end{equation}
where the physical dimension of the prefactor $D_0$ is $[D_0]=\mathrm{cm}^{2-\alpha}
\mathrm{sec}^{-1-\beta}$. The choice (\ref{eq-diff-coeff-of-x-t}) thus combines the
two parent processes (HDP and SBM) multiplicatively. In particular, we derive the
ensemble and time averaged MSDs as well as their statistical properties. We also
analyse the ageing behaviour and the effects of external confinement. We will
discuss similarities and disparities with other anomalous stochastic processes.

\section{Observables}

We start by introducing the physical observables, that we are going to analyse in
the remainder of this work. The most standard quantity to classify a stochastic
process is the MSD (\ref{msd}), which is defined in terms of the spatial average
\begin{equation}
\langle x^2(t)\rangle=\int x^2P(x,t)dx
\end{equation}
over the available space region, where $P(x,t)$ is the probability density function
to find the test particle at position $x$ at time $t$. This is typically a good
measure when many short trajectories of tracer particles are available. In many
single particle tracking experiments, however, few long time traces $x(t)$ of
length $T$ (the measurement time) are available. The time series $x(t)$ is usually
evaluated in terms of the time averaged MSD
\begin{equation}
\label{eq-TAMSD}
\overline{\delta^2(\Delta)}=\frac{1}{T-\Delta}\int_0^{T-\Delta}\Big[x(t+\Delta)-x(t)
\Big]^2dt,
\end{equation}
where $\Delta$ is the lag time. Often, the additional average
\begin{equation}
\left<\overline{\delta^2(\Delta)}\right>=\frac{1}{N}\sum_{i=1}^N\overline{\delta^2_i
(\Delta)}
\end{equation}
over $N$ individual trajectories is taken to produce smooth results.

For Brownian motion, the increments are stationary, and the time averaged MSD in
the limit of long times becomes identical to the ensemble MSD: $\lim_{T/\Delta\to
\infty}\overline{\delta^2(\Delta)}=\langle x^2(\Delta)\rangle$ \cite{pccp,pccp0,pt,
igor}. Due to this identity we call Brownian motion an ergodic process \cite{ergo}.
The same is true for anomalous diffusion processes driven by fractional Gaussian
noise \cite{goychuk,deng,pre} (although pronounced transients may become relevant
\cite{lene1,pre1}), as well as for diffusion on fractals \cite{meroz}. In contrast,
non-stationary processes are subject to the disparity
\begin{equation}
\label{web}
\lim_{T/\Delta\to\infty}\overline{\delta^2(\Delta)}\neq\langle x^2(\Delta)\rangle.
\end{equation}
This behaviour is referred to as \emph{weak ergodicity breaking\/} \cite{pccp,pccp0,
pt,igor,pnas,web,web1,web2}. For a range of anomalous diffusion processes, the
time averaged MSD in the limit $\Delta\ll T$ was found to scale \emph{linearly\/}
in the lag time,
\begin{equation}
\label{web_tamsd}
\left<\overline{\delta^2(\Delta)}\right>\simeq\frac{\Delta}{T^{1-\kappa}},
\end{equation}
despite the anomalous scaling of the ensemble MSD (\ref{msd}). Simultaneously, the
explicit dependence on the measurement time shows that the process is progressively
slowing down $(0<\kappa<1$) or picking up speed ($\kappa>1$). The specific form
(\ref{web_tamsd}) was derived, inter alia, for subdiffusive continuous time random
walks \cite{web,web1,web2}, HDPs \cite{fulinski,hdp}, and SBM \cite{fulinski,thiel,
sbm}. We note that a similar linear lag time dependence of the time averaged MSD was
also observed in ultraslow processes with a logarithmic form of the ensemble averaged
MSD \cite{anna,sinai,michael,sanders,denisov}.

Ergodic processes are reproducible in the sense that each time we evaluate some
observables for a---sufficiently long---measured time series we obtain the same
result with only minor deviations. Quantified in terms of the dimensionless
variable \cite{web2}
\begin{equation}
\xi=\frac{\overline{\delta^2}}{\left<\overline{\delta^2}\right>},
\end{equation}
this means that the associated distribution $\phi(\xi)$ converges to the delta
function, $\lim_{T/\Delta\to\infty}\phi(\xi)=\delta(\xi-1)$ \cite{pccp,pccp0,pt,
web2}. Weakly non-ergodic processes have different limit distributions for $\phi(
\xi)$. For instance, subdiffusive continuous time random walks have a finite value
$\phi(\xi=0)$ and a distinct contribution away from the ergodic value $\xi=1$
\cite{pccp,pccp0,pt,web2}. HDPs have shapes of $\phi(\xi)$ that are similar to
Gamma distributions with $\phi(\xi=0)=0$ \cite{hdp}. SBMs are weakly non-ergodic
but asymptotically reproducible \cite{thiel,sbm}.

The spread of the time averaged MSD at given lag time $\Delta\ll T$ is quantified
by the ergodicity breaking parameter \cite{pccp,pccp0,pt,web2}
\begin{equation}
\label{eq-eb-via-xi}
\mathrm{EB}(\Delta)=\left<\xi^2(\Delta)\right>-\left<\xi(\Delta)\right>^2=
\left<\xi^2(\Delta)\right>-1.
\end{equation}
For Brownian motion EB vanishes linearly with $\Delta/T$ \cite{pccp},
\begin{equation}
\lim_{\Delta/T\to0}\mathrm{EB}_{\mathrm{BM}}(\Delta)=\frac{4\Delta}{3T}. 
\label{eq-eb-bm}
\end{equation} 
Sometimes, the alternative ergodic parameter
\begin{equation}
\mathcal{EB}(\Delta)=\frac{\left<\overline{\delta^2(\Delta)}\right>}{\left<x^2(
\Delta)\right>}
\label{eq-eb2}
\end{equation}
is invoked to provide some additional information about the ergodic 
properties of the process \cite{pccp,aljash13}.

The ergodicity breaking parameter (\ref{eq-eb-via-xi}) for any finite ratio
$\Delta/T$ for non-ergodic processes is
larger than the corresponding Brownian value (\ref{eq-eb-bm}) and may not vanish in
the limit of long measurement times. As we show below the ergodicity breaking
parameter (\ref{eq-eb-via-xi}) is a sensitive measure for the physical processes
generating a given anomalous diffusion process. Other observables such as the
ensemble and time averaged MSDs
often feature similar or even identical functional forms for different processes
and are thus not suitable to discern a specific process, compare the discussion in
Refs.~\cite{pccp,pvar,nctrw}. 

Finally, we address the ageing behaviour of the anomalous diffusion processes. This
is the explicit dependence of physical observables on the ageing time $t_a$ elapsing
between the initiation of the process at $t=0$ and the start of the measurement at
$t_a$. In particular, for the time averaged MSD of the aged system the evaluation
of the associated time series is then performed in terms of
\begin{equation}
\label{eq-TAMSD-aged}
\overline{\delta^2(\Delta;t_a)}=\frac{1}{T-\Delta}\int\limits_{t_a}^{t_a+T-\Delta}
\Big[x(t+\Delta)-x(t)\Big]^2dt.
\end{equation} 
For subdiffusive continuous time random walk processes \cite{johannes}, HDPs
\cite{hdp3}, and SBM \cite{hadiseh}, it was found that the aged and non-aged time
averaged MSDs for $\Delta\ll T$ fulfil the relation
\begin{equation}
\label{age_tamsd}
\left<\overline{\delta^2(\Delta;t_a)}\right>=\Lambda_{\kappa}\left(\frac{t_a}{T}
\right)\left<\overline{\delta^2(\Delta;0)}\right>
\end{equation}
and thus differ only by the ageing factor
\begin{equation}
\label{age_dep}
\Lambda_{\kappa}\left(\frac{t_a}{T}\right)\sim\left(1+\frac{t_a}{T}\right)^{\kappa}
-\left(\frac{t_a}{T}\right)^{\kappa}
\end{equation}
containing, multiplicatively, all information on the ageing and measurement times.

In what follows, we analyse these quantities in detail for the GDP composed of the
HDP and SBM parent processes.

\section{Generalised diffusion processes}

Based on the spatio-temporal dependence (\ref{eq-diff-coeff-of-x-t}) of the
diffusion coefficient we define the generalised diffusion process in terms of
the overdamped Langevin equation
\begin{equation}
\frac{dx(t)}{dt}=\sqrt{D(x,t)}\times\xi(t).
\label{eq-dc}
\end{equation}
The noise $\xi(t)$ is Gaussian with unit variance and zero mean \cite{hdp,sbm}.
For the position coordinate $x(t)$, this is a multiplicative equation, which we
extensively tested for HDPs \cite{hdp,hdp1,hdp2,hdp3}. With respect to the time
dependence of $D(x,t)$, we analysed this Langevin equation in the context of
SBM processes \cite{sbm}.

We interpret the Langevin equation in the Stratonovich sense. In the mid-point
discrete version, the particle displacement in the $(i+1)$st step thus becomes
\cite{hdp}
\begin{equation}
\label{eq-simul-scheme}
x_{i+1}-x_i=\sqrt{2\left[D\left(\frac{x_{i+1}+x_i}{2},t_i\right)+D_{\mathrm{off}}
\right]}\times(y_{i+1}-y_i),
\end{equation}
where the increments $(y_{i+1}-y_i)$ of the Wiener process represent a
$\delta$-correlated Gaussian noise with unit variance. Unit time intervals
separate the consecutive iteration steps. To avoid a trapping of particles in the
regions of zero diffusivity for certain exponents for $\alpha$ and $\beta$, the 
form (\ref{eq-simul-scheme}) chosen for the simulations was regularised by addition
of the small constant $D_{\mathrm{off}}=10^{-3}$, compare Refs.~\cite{hdp,hdp2,hdp3}
for more details on the simulation procedure. The particle's initial starting
position in all the results presented below is $x_0=0.1$ (see
Ref.~\cite{hdp3} for details of the effects of $x_0$). The discrete scheme
(\ref{eq-simul-scheme}) is supplied with either natural or reflective boundary
conditions in the following.

From the Langevin equation (\ref{eq-dc}) it is straightforward to derive the
diffusion equation (compare Ref.~\cite{hdp})
\begin{equation}
\label{diffeq}
\frac{\partial}{\partial t}P(x,t)=\frac{\partial}{\partial x}\left[\sqrt{D(x,t)}
\frac{\partial}{\partial x}\left(\sqrt{D(x,t)}P(x,t)\right)\right].
\end{equation}
Following the Stratonovich interpretation we see that the diffusion coefficient
appears symmetric with respect to the Laplacian operator. In the limits $\beta
\to0$ and $\alpha\to0$, respectively, we recover the diffusion equations with
purely position dependent \cite{hdp} and time dependent \cite{sbm} diffusivity.

\subsection{Unconfined motion}

\begin{figure}
\begin{center}
\includegraphics[height=6.4cm]{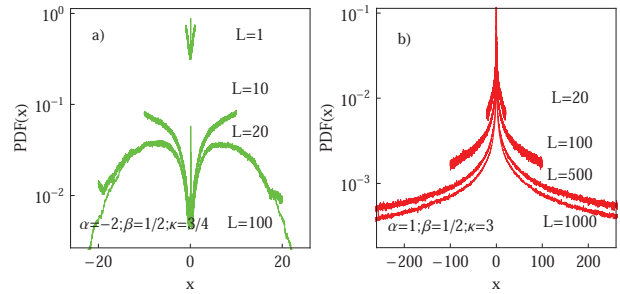}
\end{center}
\caption{Probability density function $P(x,t)$  for free and confined (see below)
GDPs for (a) subdiffusive and (b) superdiffusive choices of the critical exponents
$\alpha$ and $\beta$, as indicated in the plots. For subdiffusive GDPs the small
spike at $x=0.1$ is a remainder of the initial starting position of the particles. 
For superdiffusive GDPs this is not visible due to the fast formation of the peak
of $P(x,t)$ at $x=0$ in the region of the slowest diffusivity. Parameters: $T=10^4$,
$N=10^3$. The $L$ values are as indicated.}
\label{fig-free-ghdps-pdf} 
\end{figure}

For unconfined motion we apply the natural boundary conditions $\lim_{|x|\to\infty}
P(x,t)=0$ to the GDP diffusion equation (\ref{diffeq}) to find the probability
density function
\begin{equation}
\label{pdf}
P(x,t)=\frac{|x|^{1/p-1}}{\sqrt{4\pi D_0t^{1+\beta}}}\exp\left(-\frac{|x|^{2/p}}{
(2/p)^2D_0t^{1+\beta}}\right),
\end{equation}
where we used the abbreviation
\begin{equation}
p=\frac{2}{2-\alpha}.
\end{equation}
The functional form of the probability density function (\ref{pdf}) thus combines
the spatial features of the HDP process with the space dependence $\simeq|x|^{
\alpha}$ of the diffusivity of SBM \cite{hdp} and the modified time dependence due to
the contribution $\simeq t^{\beta}$ of the diffusivity \cite{sbm}. In particular,
for $p>1$ ($p<1$) we observe a stretched (compressed) Gaussian shape for larger
$x$ as well as a cusp (dip to zero) close to the origin. This behaviour is
supported from simulations in Fig.~\ref{fig-free-ghdps-pdf}.

\begin{figure}
\begin{center}
\includegraphics[width=8cm]{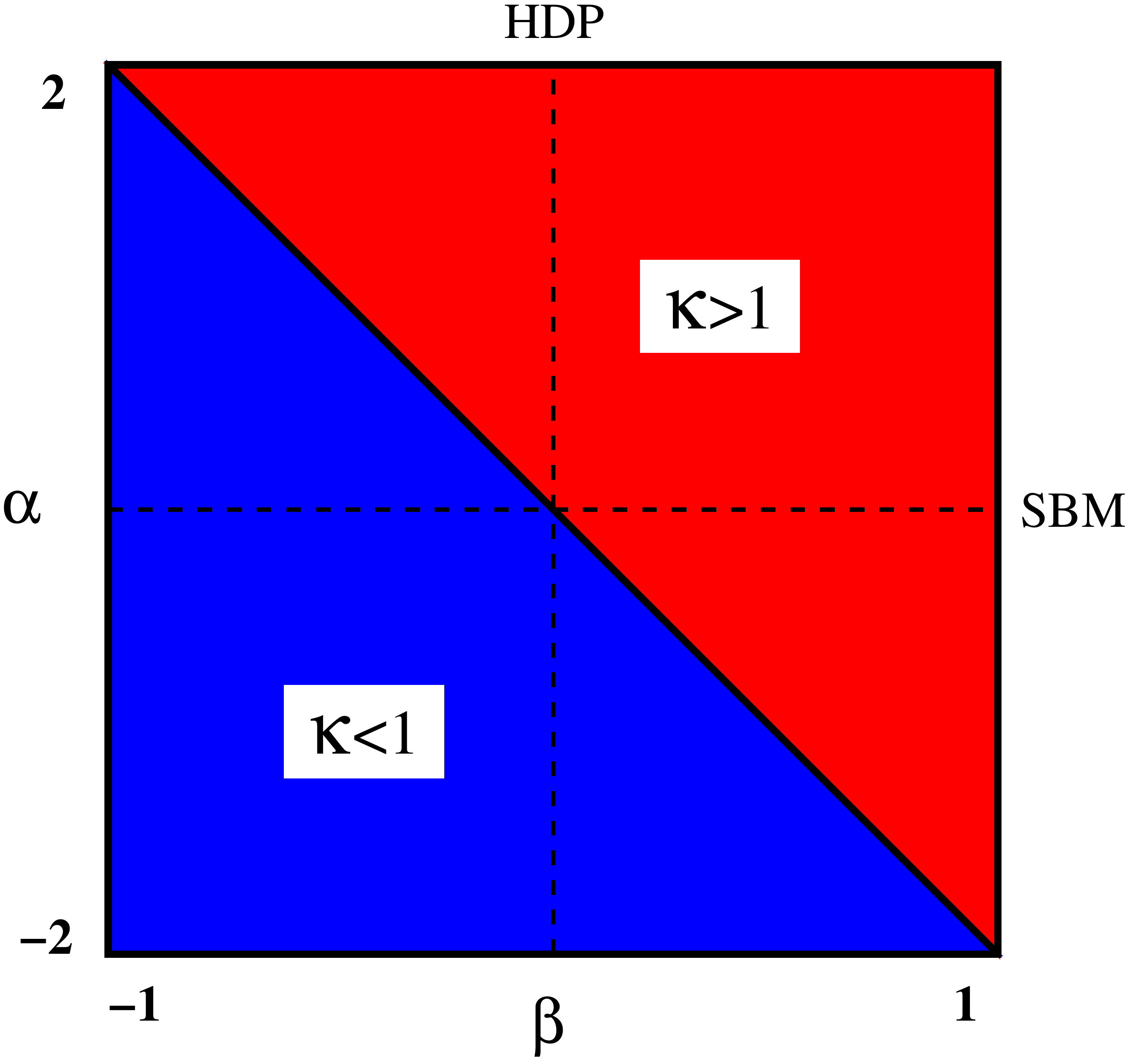}
\end{center}
\caption{Phase space of the generalised diffusion process. In the red (blue) areas
the GDP is superdiffusive (subdiffusive). The dashed lines at $\alpha=0$ ($\beta=0$)
represent pure SBM (HDP), respectively.}
\label{phase}
\end{figure}

\begin{figure}
\includegraphics[width=14cm]{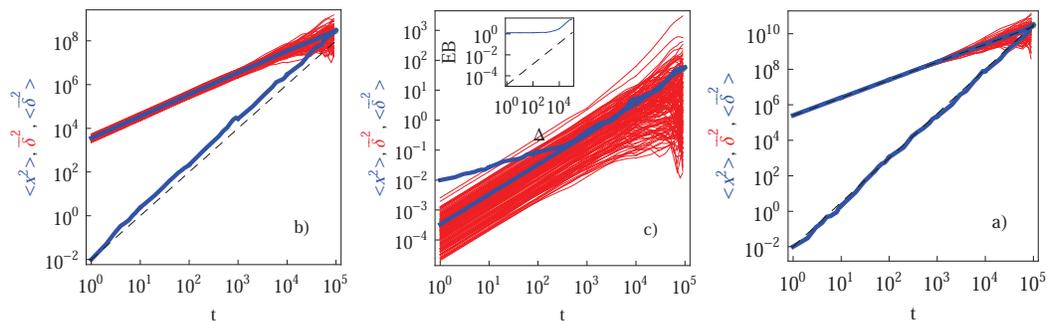}
\caption{a) Ensemble and time averaged MSDs for unconfined GDPs for $\alpha=-1/2$,
$\beta=3/2$, $T=10^5$, and $N=100$.
b) Ensemble and time averaged MSDs of
unconfined GDPs with $\alpha=1$ and $\beta=-1/2$, that is, for a linear
time dependence of the ensemble MSD ($\kappa=1$). Note the initial relaxation
of the ensemble MSD to the asymptotic behaviour $\langle x^2(t)\rangle\sim t^1$.
The inset shows the behaviour of the ergodicity breaking parameter with the
Brownian asymptote (\ref{eq-eb-bm}).
c) Ensemble and time averaged MSDs for unconfined SBM process for
$\beta=3/2$. The limiting laws (\ref{gdp_msd}) and (\ref{gdp_tamsd}) with
$p=1$ are represented by the dashed curves. Parameters: $T=10^5$, $N=50$.}
\label{fig-sbm-free-MSD}
\end{figure}

Integration
based on Eq.~(\ref{pdf}) produces the ensemble averaged MSD
\begin{equation}
\label{gdp_msd}
\langle x^2(t)\rangle=\frac{\Gamma(1/2+p)}{\pi^{1/2}}\left(\frac{2}{p}\right)^{2p}
\left(D_0t^{1+\beta}\right)^p.
\end{equation}
The scaling exponent
\begin{equation}
\kappa=(1+\beta)p
\end{equation}
is thus given by the product of the HDP and SBM exponents. In the limits $\beta=0$
and $\alpha=0$ we recover the scaling exponents $\kappa=p$ and $\kappa=1+\beta$ of
the pure parent processes for HDP and SBM, respectively \cite{fulinski,hdp,sbm}. For
the parameter range $-2<\alpha<2$ and $-1<\beta<1$ of the HDP and SBM processes,
the phase space of the GDP is depicted in Fig.~\ref{phase}: for all parameter
values in the red (blue) areas, the GDP is superdiffusive (subdiffusive). Note that
we also simulated the GDP for the temporal exponent $\beta=3/2$ outside of the
typical SBM range $-1<\beta<1$ in Fig.~\ref{fig-sbm-free-MSD}a, observing good
agreement with the prediction for the MSDs (\ref{gdp_msd}) and (\ref{gdp_tamsd}).

In analogy to the derivations in Refs.~\cite{hdp,sbm}, we obtain the time averaged
MSD in the limit $\Delta\ll T$,
\begin{eqnarray}
\nonumber
\left<\overline{\delta^2(\Delta)}\right>&\sim&\frac{\Gamma(1/2+p)}{\pi^{1/2}}\left(
\frac{2}{p}\right)^{2p}D_0^p\frac{\Delta}{T^{1-(1+\beta)p}}\\
&\sim&\langle x^2(\Delta)\rangle\left(\frac{\Delta}{T}\right)^{1-(1+\beta)p}.
\label{gdp_tamsd}
\end{eqnarray}
Time and ensemble averages are thus disparate, and in analogy to the parent
processes HDP and SBM we obtain weak ergodicity breaking. The alternative
ergodicity breaking parameter for the GDP thus becomes
\begin{equation}
\mathcal{EB}(\Delta)\sim\left(\frac{\Delta}{T}\right)^{1-(1+\beta)p}.
\label{gdp_alteb}
\end{equation}
The explicit dependence on the measurement time $T$ in relations (\ref{gdp_tamsd})
and (\ref{gdp_alteb}) is a signature of the non-stationary character of the GDP.
As function of $T$ the amplitude of the time averaged MSD continuously decreases
(increases) for subdiffusive (superdiffusive) parameters.

\begin{figure}
\begin{center}
\includegraphics[width=8cm]{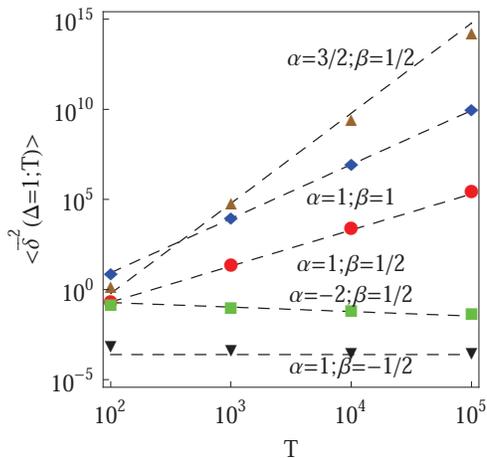}
\end{center}
\caption{Time averaged MSD for GDPs in the limit $\Delta/T\ll1$ ($\Delta=1$). 
The asymptotes are given by Eq.~(\ref{gdp_tamsd}) and the values of the exponents
$\alpha$ and $\beta$ are as indicated. Good agreement with the predicted scaling
behaviour is observed.}
\label{fig-GDP-free-TAMSD-versus-T}
\end{figure}

Fig.~\ref{fig-GDP-free-TAMSD-versus-T} shows the time averaged MSD from simulations
for a range of values of the scaling exponents $\alpha$ and $\beta$. The predicted
scaling $\langle\overline{\delta^2}\rangle\simeq T^{(1+\beta)p-1}$ according to
Eq.~(\ref{gdp_tamsd}) agrees nicely with the simulated data. The linear lag time
dependence $\langle\overline{\delta^2}\rangle\simeq\Delta$ is supported by the
simulations shown in Fig.~\ref{fig-sbm-free-MSD}a and below in
Fig.~\ref{fig-ghdps-confined-MSD}.

Note that for those combinations of the scaling exponents $\alpha$ and $\beta$ for
which the ensemble MSD grows linearly ($\kappa=1$, see the diagonal separating
subdiffusion and superdiffusion in Fig.~\ref{phase}) the GDP is still weakly
non-ergodic, as illustrated by the irreproducibility of the individual time averaged
MSDs and their inequivalence with the ensemble MSD shown in
Fig.~\ref{fig-sbm-free-MSD}b for $\alpha=1$ and $\beta=-1/2$. Despite the
fact that the trends of the spatial and temporal variation of the GDP diffusivity
compensate each other in terms of the MSD the overall process is still
non-stationary.

\begin{figure}
\begin{center}
\includegraphics[width=14cm]{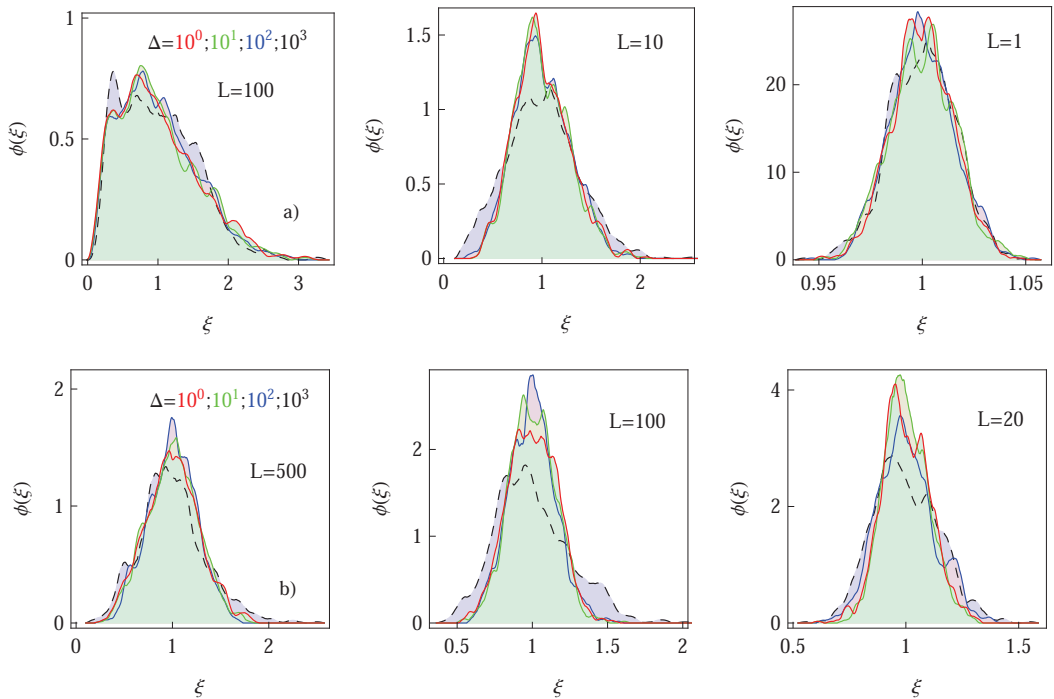}
\end{center}
\caption{Amplitude scatter distribution $\phi(\xi)$ of individual time averaged MSD
traces of the GDP for varying lag times $\Delta$ and the extent of confinement. 
As parameters we chose $T=10^4$, $N=10^3$, as well as for the top row $\alpha=-2$
and $\beta=1/2$ (subdiffusion), for the bottom row $\alpha=1$, $\beta=1/2$
(superdiffusion). Different degrees of confinement were imposed, the left panel
in each row corresponds to effectively free motion.}
\label{fig-free-ghdps-tamsd-scatter} 
\end{figure}

The scatter of the time averaged MSD from individual simulated trajectories is a
sensitive function of the diffusivity $D(x,t)$, as shown in
Fig.~\ref{fig-free-ghdps-tamsd-scatter}. The spread of $\overline{\delta^2}$ 
becomes broader as the value of the spatial exponent $\alpha$ approaches its
critical value $\alpha\to2$. This is a characteristic property of HDPs \cite{hdp3}.  
With increasing $\Delta$ the spread of $\overline{\delta^2}$ traces increases only
slightly. For subdiffusive GDPs the distribution $\phi(\xi)$ is similar to the
asymmetric Rayleigh form discussed in Ref.~\cite{hdp}.

\begin{figure}
\includegraphics[width=14cm]{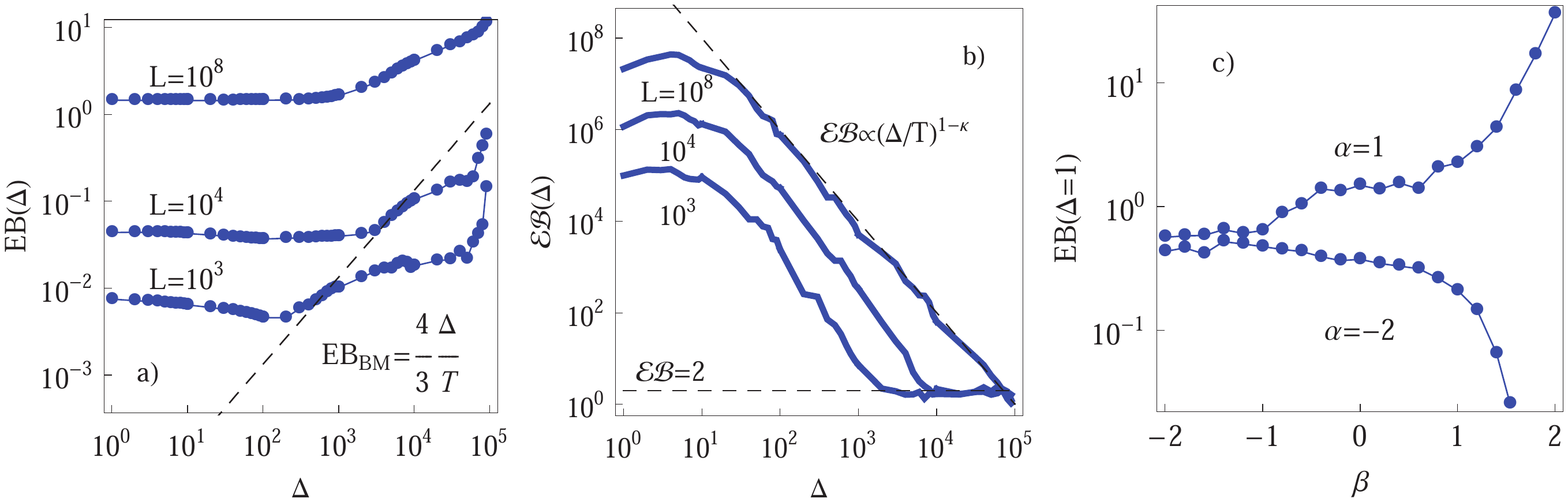}
\caption{Ergodicity breaking parameters $\mathrm{EB}$ (panel a) and $\mathcal{EB}$
(panel b) for free ($L=10^8$) and confined superdiffusive GDPs ($\beta=1$,
$\alpha=1/2$). The asymptotes (\ref{eq-eb-bm}) for Brownian motion in (a) and
(\ref{gdp_alteb}) in (b) are shown as the dashed lines. The values of the
confinement $L$ are as indicated. Parameters: $T=10^5$, $N=150$.
Panel c): log-linear plot of $\mathrm{EB}$ at $\Delta=1$ for different $\beta$
values for free GDPs. Parameters: $T=10^4$, $N=500$.}
\label{fig-ghdps-confined-EB2} 
\end{figure}

The ergodicity breaking parameter $\mathrm{EB}(\Delta)$ of GDPs is illustrated
in Fig.~\ref{fig-ghdps-confined-EB2}a. The deviations from the Brownian result
(\ref{eq-eb-bm}) in the limit $\Delta/T\ll1$ quantifies the departure of the
diffusing particles from the ergodic behaviour. We find that for GDPs the value
of the ergodicity breaking parameter does not vanish in the limit $\Delta/T\ll1$,
see Fig.~\ref{fig-ghdps-confined-EB2}a. For GDPs the value of $\mathrm{EB}$ in
the limit $\Delta/T\to0$ varies significantly with $\beta$, as shown in
Fig.~\ref{fig-ghdps-confined-EB2}c. We note that the diffusing particles under
lesser confinement produce larger values of $\mathrm{EB}$, compare the curves in
Fig.~\ref{fig-ghdps-confined-EB2}a, see below. We also note that varying initial
conditions $x_0$ alter the values of $\mathrm{EB}$ (not shown). The auxiliary
ergodicity breaking parameter for free and confined GDP motion is shown in
Fig.~\ref{fig-ghdps-confined-EB2}b.

\subsection{Ageing effects}
\label{sec-aged-GDP}

\begin{figure}
\includegraphics[width=16cm]{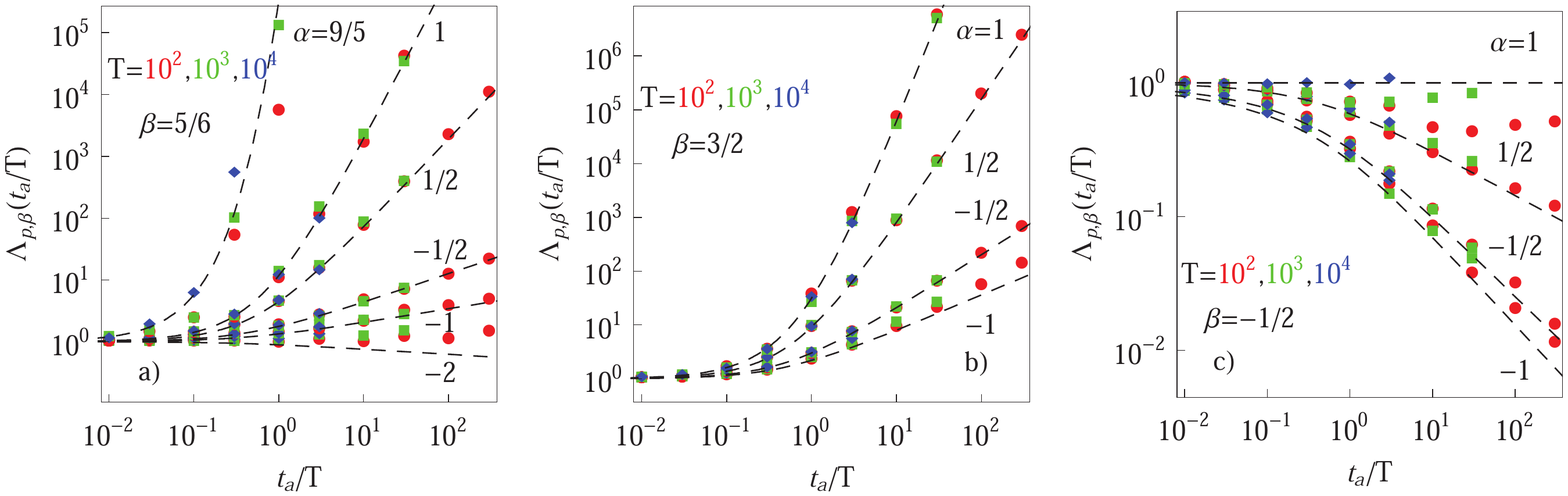}
\caption{Ageing factor $\Lambda_{p,\beta}(t_a/T)$ for aged GDPs in the
limit $\Delta/T\ll1$ ($\Delta=1$). The values of $\alpha$ are given in the plots.
We show the values $\beta=5/6$ (panel a), $\beta=3/2$ (panel b), and $\beta=-1/2$
(panel c). The asymptotes (\ref{eq-lambda-aged-ghdp}) are represented by the dashed
lines. Parameters: the overall length of the traces is $10^5$ steps, of which the
evaluated trace lengths $T$ are indicated in the plots. We averaged over $N=200$
traces for each data point.}
\label{fig-aged-ghdps-LAMBDA} 
\end{figure}

The effect of ageing, that is, the dependence of physical observables on the
time difference $t_a$ between system initiation at $t=0$ and start of the
measurement of the process in analogy to the results for HDPs \cite{hdp3} and
SBM \cite{hadiseh} factorises according to Eq.~(\ref{age_tamsd}). The ageing
factor (\ref{age_dep})
\begin{equation}
\label{eq-lambda-aged-ghdp}
\Lambda_{p,\beta}\left(\frac{t_a}{T}\right)\sim\left(1+\frac{t_a}{T}\right)^{p(1+
\beta)}-\left(\frac{t_a}{T}\right)^{p(1+\beta)}
\end{equation}
now has a parametric dependence on both $p=2/(2-\alpha)$ and $\beta$. This
scaling with the ageing time is indeed supported by our computer simulations
for a number of combinations of the scaling exponents $\alpha$ and $\beta)$,
as shown in Fig.~\ref{fig-aged-ghdps-LAMBDA}. Fig.~\ref{fig-aged-ghdps-LAMBDA}b
shows that for aged GDPs the scaling function (\ref{eq-lambda-aged-ghdp}) remains
valid also for $\beta>1$, see the discussion for ageing SBM below. Note that for
subdiffusive GDPs in general longer traces are needed to obtain the same degree
of convergence to Eq.~(\ref{eq-lambda-aged-ghdp}), as detailed in
Fig.~\ref{fig-aged-ghdps-LAMBDA}c.

\subsubsection*{Ageing SBM.}
\label{sec-aged-SBM}

\begin{figure}
\begin{center}
\includegraphics[width=8cm]{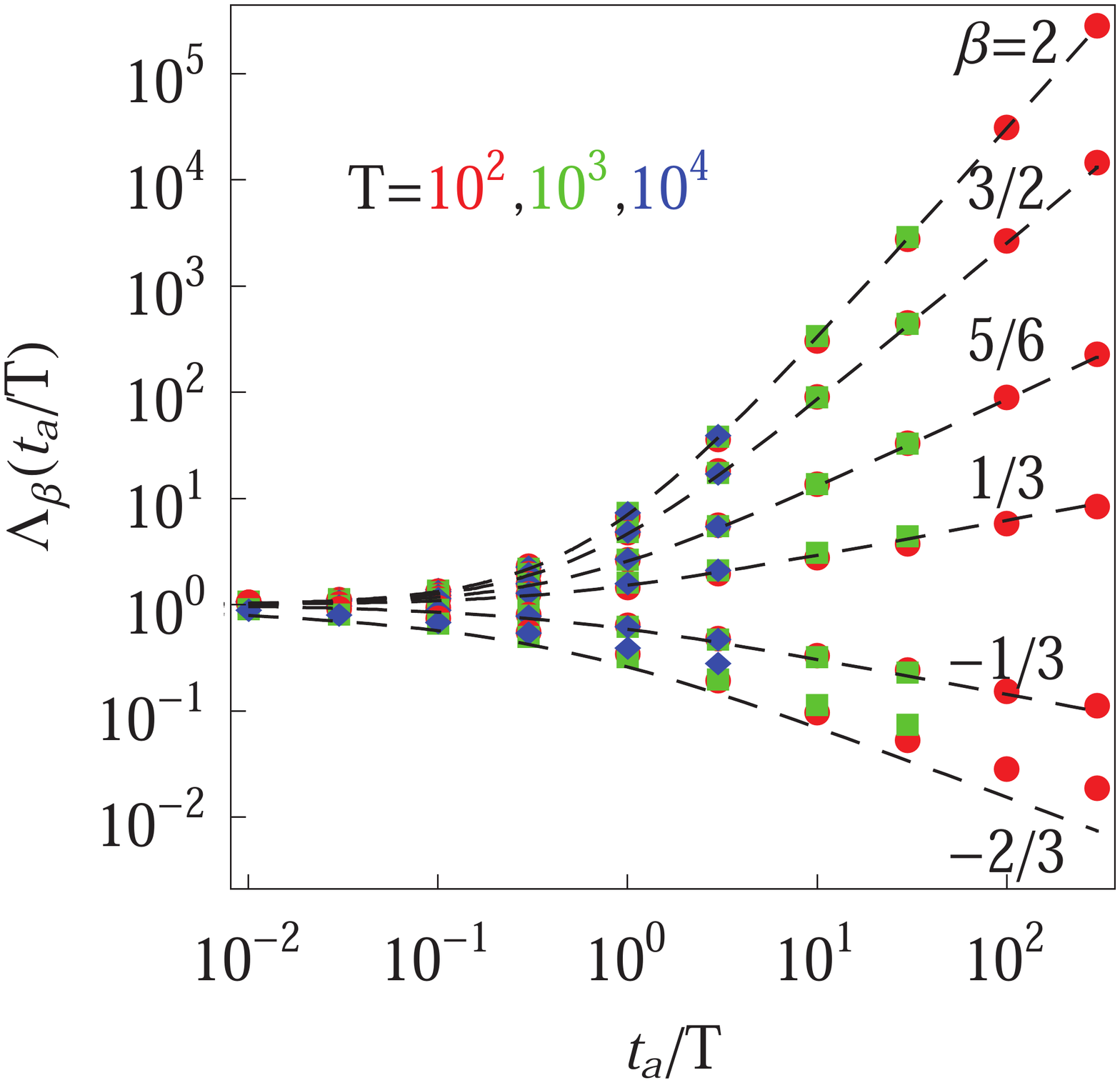}
\end{center}
\caption{Ageing factor $\Lambda_{\beta}$ for ageing SBM, the dashed lines
represent the asymptote (\ref{eq-lambda-aged-ghdp}) for $p=1$. Parameters: the total
trajectory length is $10^5$ steps and the length $T$ evaluated in the graphs is
stated in the plot panels. The averaging was over $N\approx100$ trajectories for
each $\beta$ value, and we chose $\Delta=1$.}
\label{fig-sbm-aged}
\end{figure}

Let us briefly digress to study the ageing behaviour of pure SBM. From the results
of Ref.~\cite{sbm} together with the definition of the ageing time averaged MSD
(\ref{eq-TAMSD-aged}) we find the full form of the ageing factor
\begin{eqnarray}
\label{eq-lambda-aged-sbm}
\hspace*{-2.48cm}
\Lambda_\beta(t_a/T)=\frac{(1+t_a/T)^{\beta+2}+(t_a/T)^{\beta+2}-(t_a/T+\Delta/T)^{
\beta+2}-(1+t_a/T-\Delta/T)^{\beta+2}}{1-(\Delta/T)^{\beta+2}-(1-\Delta/T)^{\beta+2}}
\end{eqnarray}
that for short ageing $t_a/T\ll1$ and lag times $\Delta/T\ll1$ reduces to relation
(\ref{eq-lambda-aged-ghdp}) with $p=1$. This implies that for superdiffusive
SBM with $\beta>0$ the magnitude of time averaged MSD increases with the ageing time
$t_a/T$ and decreases for subdiffusive SBM with $\beta<0$, as shown in
Fig.~\ref{fig-sbm-aged}. Physically this is a consequence of the continued
acceleration of particles in the superdiffusive case, contrasting the progressively
localised particles for subdiffusion. We refer to Ref.~\cite{hadiseh} for 
more details on the properties of ageing SBM including the mean first passage time
density.

\begin{figure*}
\includegraphics[width=16cm]{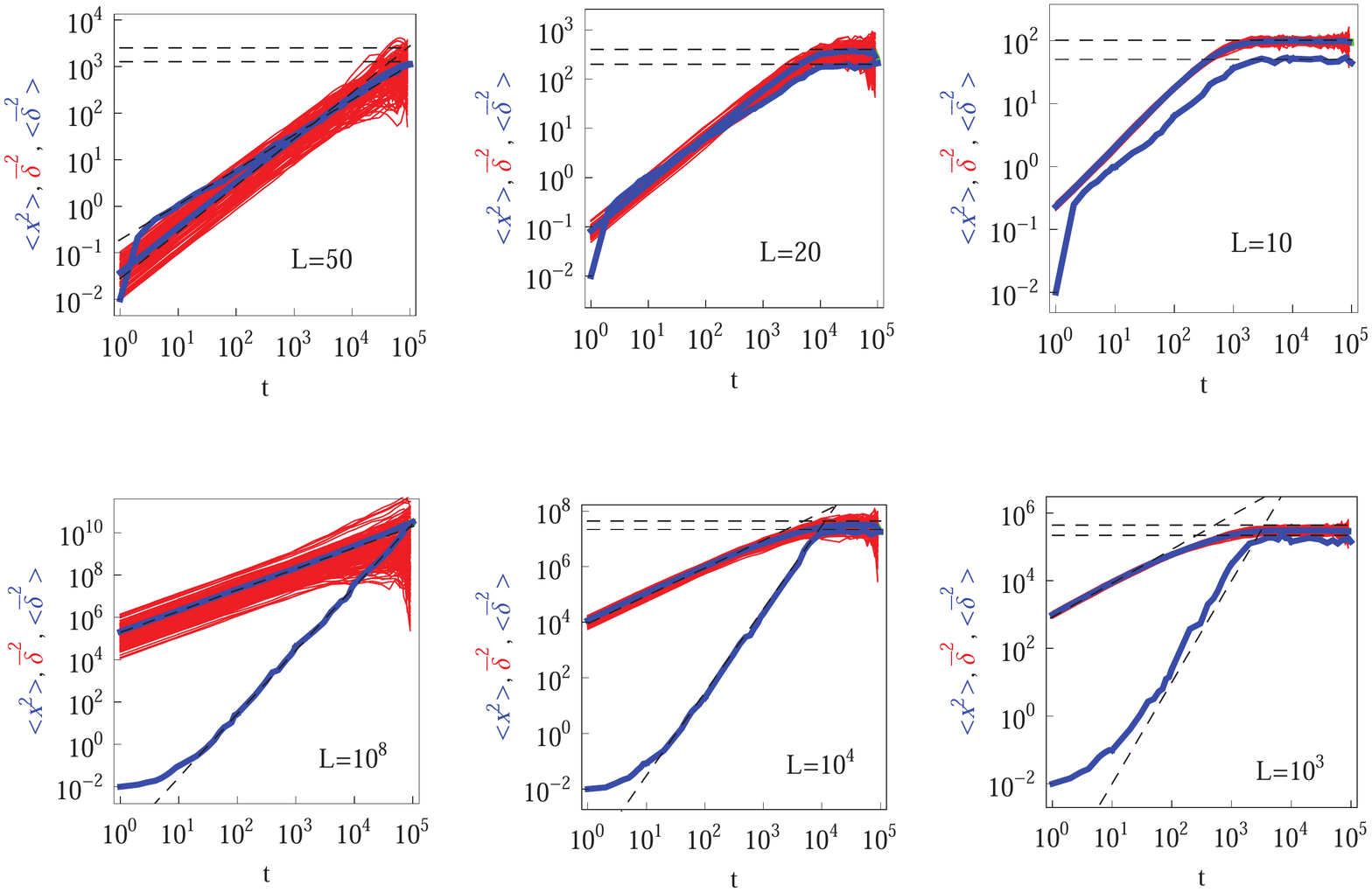}
\caption{Ensemble and time averaged MSDs for confined subdiffusive ($\alpha=-2$
and $\beta=1/2$, top row) and superdiffusive ($\beta=1$ and $\alpha=1/2$, bottom
row) GDPs. The limiting behaviours (\ref{gdp_msd}) and (\ref{gdp_tamsd}) for
unconfined motion as well as the plateau value (\ref{eq-confined-ghdp-msd}) under
confinement are indicated by the dashed curves. Parameters: $T=10^5$ and $N=150$.
The width of the confining interval $L$ is indicated in the panels, the value
$L=10^8$ corresponds to an effectively unconfined situation.}
\label{fig-ghdps-confined-MSD} 
\label{fig-ghdps-confined-MSD-subdiffusive} 
\end{figure*}

We support the limiting form (\ref{eq-lambda-aged-ghdp}) with $p=1$ of the ageing
factor for ageing SBM from computer simulations in
Fig.~\ref{fig-sbm-aged} for a range of scaling exponents $\beta$. The agreement of
the simulations with the asymptote (\ref{eq-lambda-aged-ghdp}) is particularly good
for the case of superdiffusive SBM with $T=10^5$ steps. Strongly subdiffusive SBMs 
likely require longer traces to converge better to (\ref{eq-lambda-aged-ghdp}).
Performing simulations for the range $\beta>1$ we also reveal excellent agreement
with the ageing factor (\ref{eq-lambda-aged-ghdp}), see the top curves in
Fig.~\ref{fig-sbm-aged}.

\subsection{Confined motion}
\label{sec-confined-GDP}

Let us now address GDPs in the confinement of hard walls. To this end we establish
reflecting boundary conditions at $x=\pm L$. Fig.~\ref{fig-ghdps-confined-MSD}
demonstrates that for weak confinement (large $L$ values) and at shorter (lag)
times the ensemble and time averaged MSDs behave similarly to those of the
unconfined process, as expected. At stronger confinement and at longer (lag) times
the MSD saturates at the plateau value
\begin{equation}
\left<x^2\right>_{\mathrm{st}}\sim p^{-3/5}L^2/3.
\label{eq-confined-ghdp-msd}
\end{equation}
The $\alpha$-dependence of the value (\ref{eq-confined-ghdp-msd}) in the prefactor
$p^{-3/5}$ reflects the fact that the associated stationary probability density
function is still position dependent due to the variation of the diffusivity, see
also Fig.~\ref{fig-free-ghdps-pdf}. For $\alpha=0$, the process assumes an
equidistribution with the value $1/(2L)$. We stress that the plateau value
(\ref{eq-confined-ghdp-msd}) is independent of the temporal scaling exponent
$\beta$. In the hard confinement by infinitely steep walls the stationary state
is generally independent of any $x$-independent noise strength. The variance
(\ref{eq-confined-ghdp-msd}) is thus geometry induced and not due to the
competition of thermal noise and confinement as, for instance, in an harmonic
confinement. In the latter case, no stationary solution is reached for SBM
\cite{sbm}.

Fig.~\ref{fig-ghdps-confined-MSD} also shows that for the same interval width $L$
subdiffusive GDPs require longer times to reach the stationary values. After
reaching the stationary plateau, the value of the time averaged MSD is twice that
of the ensemble MSD, an effect of the very definition of the time averaged MSD
\cite{pccp}. Note that due to the pole at $\Delta\to T$ in the definition
(\ref{eq-TAMSD}) of the time averaged MSD the asymptotic equivalence $\lim_{
\Delta\to T}\langle\overline{\delta^2(\Delta)}\rangle\to\langle x^2(T)\rangle$
holds \cite{pccp}.

The probability density function $P(x,t)$ of confined GDPs for different values of
the scaling exponents $\alpha$ and $\beta$ as well as of varying degrees of
confinement is shown in Fig.~\ref{fig-free-ghdps-pdf}. It is seen that when the
particle is reflected repeatedly in stronger confinement by the walls, the tails
of $P(x,t)$ become perceivably raised, to fulfil the boundary condition $\left.
\partial P(x,t)/\partial x\right|_{x=0}=0$.

The distribution of amplitudes of individual time averaged MSDs described by
$\phi(\xi)$ for confined GDPs is shown in Fig.~\ref{fig-free-ghdps-tamsd-scatter}.
The width of $\phi(\xi)$ decreases for more severe confinement, as expected.
The width of $\phi(\xi)$ for shorter measurement times $T$ is larger (not shown),
due the smaller number of reflections from the walls. For confined GDPs $\phi(\xi)$
becomes more localised and symmetric both for subdiffusive and superdiffusive
combinations of the scaling exponents $\alpha$ and $\beta$, as demonstrated in
Fig.~\ref{fig-free-ghdps-tamsd-scatter}a: as one may already venture from
Fig.~\ref{fig-ghdps-confined-MSD}, the spread of individual $\overline{\delta^2}$
decreases severely when extreme excursions of particles causing large
trajectory-to-trajectory variations are progressively impeded by the confinement.

\begin{figure}
\begin{center}
\includegraphics[width=8cm]{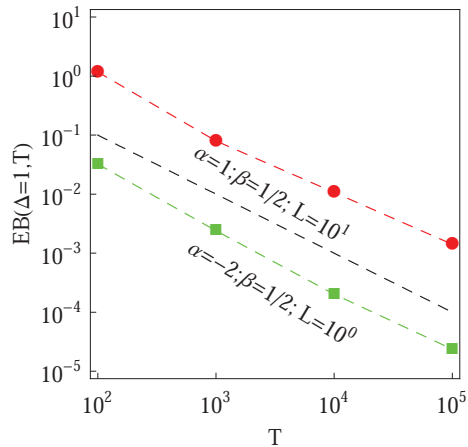}
\end{center}
\caption{Ergodicity breaking parameter $\mathrm{EB}$ for confined GDPs from
simulations. The asymptote (\ref{eq-eb-gdp-confind-versus-T}) is indicated by
the black dashed curve. Notations for the symbols are the same as in
Fig.~\ref{fig-GDP-free-TAMSD-versus-T}.}
\label{fig-ghdps-confined-EB} 
\end{figure}

The dependence of the ergodicity breaking parameter on the lag time $\Delta$ for
different interval lengths $L$ is shown in Fig.~\ref{fig-ghdps-confined-EB2}a, and
the variation of $\mathrm{EB}$ with the measurement time $T$ is illustrated in
Fig.~\ref{fig-ghdps-confined-EB}. We find for GDPs the reciprocal dependence
\begin{equation}
\mathrm{EB}(T)\sim\frac{1}{T}.
\label{eq-eb-gdp-confind-versus-T}
\end{equation}
with the trace length $T$, similar to the behaviour of confined HDPs \cite{hdp3}.
Similar to HDPs and SBM, for free and confined GDPs our simulations confirm
Eq.~(\ref{gdp_alteb})
in Fig.~\ref{fig-ghdps-confined-EB2}b. In the long time limit the
saturation plateau is reached and $\mathcal{EB}(\Delta)\to2$. We point out that
this behaviour is significantly different from that of subdiffusive continuous
time random walks, for which the power-law $\langle\overline{\delta^2(\Delta)}
\rangle\simeq(\Delta/T)^{1-\alpha}$ is obtained \cite{pnas}.

\section{Discussion and Outlook}
\label{sec-disc}

We studied the non-homogeneous and non-stationary generalised diffusion process
with the power-law diffusivity $D(x,t)\sim |x|^{\alpha}t^{\beta}$. We explored
analytically and by extensive computer simulations the competition between the
two parent processes revealing a number of universal characteristics. While the
general scaling forms for the ensemble and time averaged MSDs are similar to
those of the parent processes HDP and SBM---and in fact to those of subdiffusive
continuous time random walks \cite{pccp,pccp0,pt}---we showed that for those
combinations of the scaling exponents $\alpha$ and $\beta$ which produce a unity
anomalous diffusion exponent, $\kappa=1$, the non-stationary character of the
GDP is still visible in the irreproducibility of the time averaged MSD. For the
ageing GDP we obtained that the ageing factor has the same functional
form as for HDP, SBM, and subdiffusive continuous time random walks \cite{pccp,
hdp3,hadiseh}. We also explored the properties of GDPs confined in an interval
including the saturation of the MSD and the variation of the ergodicity breaking
parameter with the trace length and the degree of confinement. Physically, the
combination of temporal and spatial variations of the local diffusivity appears
both a natural and attractive concept to model anomalous diffusion, for instance,
in microscopic biological systems.

Restricted diffusion in various porous media was considered in terms of effective
time dependent diffusivities. Such approaches may provide some information about
the typical size of restricted regions or cavities experienced by a tracer. The
concept of time dependent diffusivities is, for instance, applied to interpret
signals in NMR experiments \cite{novikov1,novikov2,grebenkov} aimed at determining
the degree of tissue interconnectivity and permeability. Water and ion diffusion
in various human tissues is often anisotropic and sometimes anomalous \cite{nikol}.
Anisotropies in water diffusion were observed in biological tissues with a
directional/fibrous structure, for instance, in the white matter of the human brain,
nerve fibres, and muscle fibres \cite{novikov2}.

Specifically, the instantaneous diffusivity $D(t)=d\langle\mathbf{r}^2(t)\rangle/
(6dt)$ in
muscle fibres was shown to assume the long-time limit to decay $D(t)=D(\infty)+
\mathrm{const}/t^{1/2}$. Alternative power-law scaling of the form $D(t)\sim t^\mu$
corresponding to different structures of the disorder in the system were also
examined, $\mu$ often ranging in the interval $1/2<\mu<3/2$ \cite{novikov2}. The
value of $\mu=3/2$ is known for a fully uncorrelated medium in three dimensions.
In a biological context, the power $\mu$ of the tail of the diffusivity depends on
the positioning of boundaries in the system. The latter are typically cell membranes
that are only weakly permeable to water or ions. It was argued that some deceases
may caused changes in the brain or muscle tissues that can be detected by measuring
the changes in the time dependent diffusivity and the exponent $\mu$ as compared to
healthy tissues. Note that in these biological examples with time-dependent
diffusivity no complete localisation of particles in the long-time limit exists and 
the basal diffusivity $D(\infty)$ is finite.

In porous interconnected media, for instance, in a suspension of reflecting spheres
\cite{sen-3-2-tail}, the long-time diffusivity of a tracer can be defined as $\bar
D(t)=\langle\mathbf{r}^2(t)\rangle/(6t)$. It can again be represented in the time dependent form $\bar D(t)
/D_0\sim A_1+A_2/t+A_3/t^{3/2}$ \cite{loskut2012}. The leading time dependence for 
the instantaneous diffusion coefficient is then $D(t)/D_0 \sim A_1-0.5A_3t^{-3/2}$. 
Here $1/A_1$ is the tortuosity of the medium that renormalises the long-time tracer
diffusivity due to the presence of inter-connected cavities \cite{sen-3-2-tail}. 

Are such assumptions of an effective, time dependent diffusivity always justified?
In fact, any anomalous diffusion law (\ref{msd}) can be interpreted in terms of
the effective diffusivity $K_{\mathrm{eff}}(t)\simeq t^{1-\alpha}$, such that we
can rewrite Eq.~(\ref{msd}) in the form $\langle x^2(t)\rangle\simeq K_{\mathrm{eff
}}(t)t$. However, this is not sufficient to identify the observation of anomalous
diffusion with the GDP or SBM models studied here. It is important to either have
more detailed
knowledge of the underlying physical process, which may support the GDP approach
or rather other physical models such as continuous time random walks \cite{pccp,pt}.
In particular, we note that despite many close similarities, SBM and fractional
Brownian motion (FBM) capture physically strongly different situations. If direct
physical insight is not possible, the stochastic properties of the process need to
be scrutinised in more detail by using complementary diagnostic tools \cite{pccp,
pvar,vincent}.

Similar to these examples, position dependent diffusivities were explored in a
range of systems including tracer diffusion in groundwater aquifers \cite{brian}.
The position dependence is directly accessible experimentally and was systematically
sampled for the motion of smaller protein tracers in living biological cells
\cite{lang}.

We note that HDP and SBM processes may also give rise to ultraslow diffusion with
a logarithmic dependence of the MSD. For the HDP this corresponds to an exponential
variation of the diffusivity in space \cite{hdp1} while for SBM ultraslow motion
emerges in the limit $\beta\to-1$ \cite{anna}. Such ultraslow diffusion is equally
weakly non-ergodic and ageing, in analogy to diffusion in ageing environments
\cite{michael}, diffusion in quenched Sinai disorder \cite{sinai}, or many particle
systems with scale free time evolution \cite{sanders}.

\ack
We acknowledge funding from the Academy of Finland (FiDiPro scheme to RM) 
and the Deutsche Forschungsgemeinschaft (Grant CH 707/5-1 to AGC).


\section*{References}

\begin{thebibliography}{99}

\bibitem{bouchaud} J.-P. Bouchaud and A. Georges, Phys. Rep. \textbf{195}, 127
(1990).

\bibitem{report} R. Metzler and J. Klafter, Phys. Rep. \textbf{339}, 1 (2000);
J. Phys. A \textbf{37}, R161 (2004).

\bibitem{havlin} S. Havlin and D. Ben-Avraham, Adv. Phys. \textbf{51}, 187 (2002).

\bibitem{pccp} R. Metzler, J.-H. Jeon, A. G. Cherstvy, and E. Barkai, Phys. Chem.
Chem. Phys. \textbf{16}, 24128 (2014).

\bibitem{richardson} L. F. Richardson, Proc. Roy. Soc. A \textbf{110}, 709 (1926).

\bibitem{swinney} T. H. Solomon, E. R. Weeks, and H. L. Swinney, Phys. Rev. Lett.
\textbf{71}, 3975 (1993); compare the theoretical work in G. M. Zaslavsky, Phys.
Rep. \textbf{371}, 461 (2002); Hamiltonian Chaos and Fractional Dynamics (Oxford
University Press, Oxford, UK, 2005).

\bibitem{brian} R. Haggerty and S. M. Gorelick, Water Res. Res. \textbf{31},
2383 (1995); H. Scher, G. Margolin, R. Metzler, J. Klafter, and B. Berkowitz, 
Geophys. Res. Lett. \textbf{29}, 1061 (2002); B. Berkowitz, A. Cortis, M. Dentz,
and H. Scher, Rev. Geophys. \textbf{44}, RG2003 (2006).

\bibitem{scher} H. Scher and E. W. Montroll, Phys. Rev. B \textbf{12}, 2455
(1975).

\bibitem{schubert} M. Schubert et al., Phys. Rev. B \textbf{87}, 024203 (2013).

\bibitem{fcs} R. Rigler and E. S. Elson, Fluorescence Correlation Spectroscopy:
Theory and Applications (Springer, Berlin, 2011).

\bibitem{brauchle} C. Br{\"a}uchle, D. C. Lamb, and J. Michaelis, Single Particle
Tracking and Single Molecule Energy Transfer (Wiley-VCH, Weinheim, Germany,
2012); X. S. Xie, P. J. Choi, G.-W. Li, N. K. Lee, and G. Lia, Annu. Rev.
Biophys. \textbf{37}, 417 (2008).

\bibitem{wong} I. Y. Wong, M. L. Gardel, D. R. Reichman, E. R.  Weeks, M. T.
Valentine, A. R. Bausch, and D. A. Weitz, Phys. Rev. Lett. \textbf{92}, 178101
(2004).

\bibitem{mattsson} J. Mattsson, H. M. Wyss, A. Fernandez-Nieves, K. Miyazaki, Z.
B. Hu, D. R. Reichman, and D. A. Weitz, Nature \textbf{462}, 83 (2009); E. H. Zhou,
X. Trepat, C. Y. Park, G. Lenormand, M. N. Oliver, S. M. Mijalovich, C. Hardin, D.
A. Weitz, J. P. Butler, and J. J. Fredberg, Proc. Natl. Acad. Sci. \textbf{106},
10632 (2009); A. Fierro, E. del Gado, A. de Candia, and A. Coniglio,
J. Stat. Mech. \textbf{L04002} (2008).

\bibitem{weiss} J. Szymanski and M. Weiss, Phys. Rev. Lett. \textbf{103},
038102 (2009); G. Guigas, C. Kalla, and M. Weiss, Biophys. J. \textbf{93},
316 (2007); W. Pan, L. Filobelo, N. D. Q. Pham, O. Galkin, V. V. Uzunova, and
P. G. Vekilov, Phys. Rev. Lett. \textbf{102}, 058101 (2009).

\bibitem{lene1} J.-H. Jeon, N. Leijnse, L. B. Oddershede, and R. Metzler, New J.
Phys. \textbf{15}, 045011 (2013).

\bibitem{lene} D. Robert, T. H. Nguyen, F. Gallet, and C. Wilhelm, PLoS ONE
\textbf{4}, e10046 (2010); J.-H. Jeon, V. Tejedor, S. Burov, E. Barkai, C.
Selhuber-Unkel, K. Berg-S{\o}rensen, L. Oddershede, and R. Metzler, Phys. Rev.
Lett. \textbf{106}, 048103 (2011); I. Bronstein, Y. Israel, E. Kepten, S. Mai,
Y. Shav-Tal, E. Barkai, and Y. Garini, Phys. Rev. Lett. \textbf{103}, 018102 (2009);
I. Golding and E. C. Cox, Phys. Rev. Lett. \textbf{96}, 098102 (2006).

\bibitem{weigel} A. V. Weigel, B. Simon, M. M. Tamkun, and D. Krapf, Proc. Natl.
Acad. Sci. USA  \textbf{108}, 6438 (2011); S. M. A. Tabei, S. Burov, H. Y. Kim,
A. Kuznetsov, T. Huynh, J. Jureller, L. H. Philipson, A. R. Dinner, and N. F.
Scherer, Proc. Natl. Acad. Sci. USA \textbf{110}, 4911 (2013).

\bibitem{franosch} F. H\"ofling and T. Franosch, Rep. Progr. Phys. \textbf{76},
046602 (2013); M. J. Saxton and K. Jacobsen, Ann. Rev. Biophys. Biomol. Struct.
\textbf{26}, 373 (1997).

\bibitem{structure} A. Godec, M. Bauer, and R. Metzler, New J. Phys. \textbf{16},
092002 (2014); compare the experiments in C. H. Lee, A. J. Crosby, T. Emrick, and R.
C. Hayward, Macromol. \textbf{47}, 741 (2014).

\bibitem{membranes} G. R. Kneller, K. Baczynski, and M. Pasienkewicz-Gierula,
J. Chem. Phys. \textbf{135}, 141105 (2011); J.-H. Jeon, H. Martinez-Seara Monne,
M. Javanainen, and R. Metzler, Phys. Rev. Lett. \textbf{109}, 188103 (2012).

\bibitem{fractals} M. Spanner, F. H{\"o}fling, G. E. Schr{\"o}der-Turk, K. Mecke,
and T. Franosch, J. Phys. Cond. Mat. \textbf{23}, 234120 (2011); A. Klemm, R.
Metzler, and R. Kimmich, Phys. Rev. E \textbf{65}, 021112 (2002); A. Klemm and R.
Kimmich, Phys. Rev. E \textbf{55}, 4413 (1997).

\bibitem{franosch_lg} F. H{\"o}fling and T. Franosch, Phys. Rev. Lett. \textbf{98},
140601 (2007); S. Leitmann and T. Franosch, Phys. Rev. Lett. \textbf{111},
190603 (2013); E. Barkai, V. Fleurov, and J. Klafter, Phys. Rev. E \textbf{61},
1164 (2000).

\bibitem{montroll} E. W. Montroll and G. H. Weiss, J. Math. Phys. \textbf{6},
167 (1965).

\bibitem{trap} E. Bertin and J.-P. Bouchaud, Phys. Rev. E \textbf{67},
026128 (2003); C. Monthus and J.-P. Bouchaud, J. Phys. A \textbf{29}, 3847
(1996); S. Burov and E. Barkai, Phys. Rev. Lett. \textbf{98}, 250601 (2007).

\bibitem{mandelbrot} B. B. Mandelbrot and J. W. van Ness, SIAM Rev. \textbf{10},
422 (1968); compare also C. R. Acad. Sci. Paris \textbf{260}, 3274 (1965).

\bibitem{goychuk} G. Kneller, J. Chem Phys. \textbf{141}, 041105 (2014);
I. Goychuk, Phys. Rev. E \textbf{80}, 046125 (2009); Adv. Chem. Phys.
\textbf{150}, 187 (2012).

\bibitem{rouse} D. Panja, J. Stat. Mech. \textbf{L02001} (2010); \textbf{P06011}
(2010).

\bibitem{singlefile} L. Lizana, T. Ambj{\"o}rnsson, A. Taloni, E. Barkai, and M.
Lomholt, Phys. Rev. E \textbf{81}, 051118 (2010).

\bibitem{lang} T. K\"uhn et al., PLoS One \textbf{6}, e22962 (2011); B. P. English,
V. Hauryliuk, A. Sanamrad, S. Tankov, N. H. Dekker, and J. Elf, Proc. Natl. Acad.
Sci. \textbf{108}, E365 (2011).

\bibitem{fulinski} A. Fuli{\'n}ski, Phys. Rev. E \textbf{83}, 061140 (2011);
J. Chem. Phys. \textbf{138}, 021101 (2013); Acta Phys. Polonica \textbf{44},
1137 (2013).  

\bibitem{hdp} A. G. Cherstvy, A. V. Chechkin, and R. Metzler, New J. Phys.
\textbf{15}, 083039 (2013).

\bibitem{hdp1} A. G. Cherstvy and R. Metzler, Phys. Chem. Chem. Phys.
\textbf{15}, 20220 (2013).

\bibitem{hdp2} A. G. Cherstvy, A. V. Chechkin, and R. Metzler, Soft Matter
\textbf{10}, 1591 (2014).

\bibitem{hdp3} A. G. Cherstvy and R. Metzler, Phys. Rev. E \textbf{90},
012134 (2014); A. G. Cherstvy, A. V. Chechkin, and R. Metzler, J. Phys. A
\textbf{47}, 485002 (2014).

\bibitem{massignan} P. Massignan, C. Manzo, J. A. Torreno-Pina, M. F.
Garc\'{\i}a-Parako, M. Lewenstein, and G. L. Lapeyre, Jr., Phys. Rev. Lett.
\textbf{112}, 150603 (2014).

\bibitem{batchelor} G. K. Batchelor, Math. Proc. Cambridge Philos. Soc.
\textbf{48}, 345 (1952).

\bibitem{saxton1} M. J. Saxton, Biophys. J. \textbf{81}, 2226 (2001); 
G. Guigas, C. Kalla, and M. Weiss, FEBS Lett. \textbf{581}, 5094 (2007);
N. Periasmy and A. S. Verkman, Biophys. J. \textbf{75}, 557 (1998);
L. L. Latour, K. Svoboda, P. Mitra, and C. H. Sotak, Proc.
Natl. Acad. Sci. U.S.A. \textbf{91}, 1229 (1994);
J. Wu and M. Berland, Biophys. J. \textbf{95}, 2049 (2008);
J. Szymaski, A. Patkowski, J. Gapiski, A. Wilk, and R. Holyst,
J. Phys. Chem. B \textbf{110}, 7367 (2006);
J. F. Lutsko and J. P. Boon, Phys. Rev. Lett. \textbf{88}, 022108 (2013).

\bibitem{lim} S. C. Lim and S. V. Muniandy, Phys. Rev. E \textbf{66}, 021114 (2002).

\bibitem{thiel} F. Thiel and I. M. Sokolov, Phys. Rev. E \textbf{89}, 012115 (2014).

\bibitem{sbm} J.-H. Jeon, A. V. Chechkin, and R. Metzler, Phys. Chem. Chem.
Phys. \textbf{16}, 15811 (2014).

\bibitem{hadiseh} H. Safdari, A. V. Chechkin, G. R. Jafari, and R. Metzler, E-print
arXiv:1501.04810.

\bibitem{anna} A. Bodrova, A. G. Cherstvy, A. V. Chechkin, and R. Metzler
(unpublished).

\bibitem{anna1} A. Bodrova, A. V. Chechkin, A. G. Cherstvy, and R. Metzler,
E-print arXiv:1501.04173.

\bibitem{platani} M. Platani, I. Goldberg, A. I. Lamond, and J. R. Swedlow, Nature Cell Biol. \textbf{4}, 502 (2002).

\bibitem{pccp0} S. Burov, J.-H. Jeon, R. Metzler and E. Barkai, Phys. Chem.
Chem. Phys. \textbf{13}, 1800 (2011).

\bibitem{pt} E. Barkai, Y. Garini, and R. Metzler, Phys. Today
\textbf{65(8)}, 29 (2012).

\bibitem{igor} I. M. Sokolov, Soft Matter \textbf{8}, 9043 (2012).

\bibitem{ergo} L. Reichl, A Modern Course in Statistical Physics (Wiley-VCH,
Weinheim, 2009); M. Toda, R. Kubo, and N. Sait{\'o}, Statistical Physics I:
Equilibrium Statistical Mechanics (Springer, Heidelberg, 1992); A. Y. Khinchin,
Mathematical foundations of statistical mechanics (Dover Publications Inc., New
York, NY, 2003).

\bibitem{deng} W. Deng and E. Barkai, Phys. Rev. E \textbf {79}, 011112 (2009).

\bibitem{pre} J.-H. Jeon and R. Metzler, J. Phys. A \textbf{43}, 252001 (2010);
Phys. Rev. E \textbf{85}, 021147 (2012).

\bibitem{pre1} J.-H. Jeon and R. Metzler, Phys. Rev. E \textbf{85}, 021147 (2012);
J. Kursawe, J. H. P. Schulz, and R. Metzler, Phys. Rev. E \textbf{88}, 062124 (2013).

\bibitem{meroz} Y. Meroz, I. M. Sokolov, and J. Klafter, Phys. Rev. E
\textbf{81}, 010101(R) (2010).

\bibitem{pnas} S. Burov, R. Metzler and E. Barkai, 
Proc. Natl. Acad. Sci. U. S. A. \textbf{107}, 13228 (2010).

\bibitem{web} J.-P. Bouchaud, J. Phys. (Paris) I \textbf{2}, 1705 (1992); G.
Bel and E. Barkai, Phys. Rev. Lett. \textbf{94}, 240602 (2005);
G. Bel and E. Barkai, Phys. Rev. E \textbf{73}, 016125 (2006);
A. Rebenshtok
and E. Barkai, Phys. Rev. Lett. \textbf{99}, 210601 (2007); M. A. Lomholt,
I. M. Zaid, and R. Metzler, Phys. Rev. Lett. \textbf{98}, 200603 (2007); G.
Aquino, P. Grigolini, and B. J. West, Europhys. Lett. \textbf{80}, 10002 (2007).

\bibitem{web1} A. Lubelski, I. M. Sokolov, and J. Klafter, Phys. Rev. Lett.
\textbf{100}, 250602 (2008); I. M. Sokolov, E. Heinsalu, P. H{\"a}nggi, and I.
Goychuk, Europhys. Lett. \textbf{86}, 041119 (2010); M. Khoury, A. M. Lacasta,
J. M. Sancho, and K. Lindenberg, Phys. Rev. Lett. \textbf{106}, 090602 (2011);
M. J. Skaug, A. M. Lacasta, L. Ramirez-Piscina, J. M. Sancho, K. Lindenberg, and
D. K. Schwartz, Soft Matter \textbf{10}, 753 (2014).

\bibitem{web2} Y. He, S. Burov, R. Metzler, and E. Barkai, Phys. Rev. Lett.
\textbf{101}, 058101 (2008).

\bibitem{sinai} A. Godec, A. V. Chechkin, E. Barkai, H. Kantz, and R. Metzler,
J. Phys. A \textbf{47}, 492002 (2014).

\bibitem{michael} M. A. Lomholt, L. Lizana, R. Metzler, and T. Ambj{\"o}rnsson,
Phys. Rev. Lett. \textbf{110}, 208301 (2013).

\bibitem{sanders} L. P. Sanders, M. A. Lomholt, L. Lizana, K. Fogelmark, R. Metzler,
and T. Ambj{\"o}rnsson, New J. Phys. \textbf{16}, 113050 (2014).

\bibitem{denisov} S. I. Denisov, S. B. Yuste, Yu. S. Bystrik, H. Kantz, and K.
Lindenberg, Phys. Rev. E \textbf{84}, 061143 (2011).

\bibitem{aljash13} A. Godec and R. Metzler, Phys. Rev. Lett. \textbf{110}, 020603
(2013).

\bibitem{pvar} M. Magdziarz, A. Weron, K. Burnecki and J. Klafter, Phys. Rev. Lett.
\textbf{103}, 180602 (2009).

\bibitem{nctrw} J.-H. Jeon, E. Barkai and R. Metzler, J. Chem. Phys. \textbf{139},
121916 (2013).

\bibitem{johannes} J. H. P. Schulz, E. Barkai, and R. Metzler, Phys. Rev. Lett.
\textbf{110}, 020602 (2013); Phys. Rev. X \textbf{4}, 011028 (2014).

\bibitem{novikov1} D. S. Novikov, E. Fieremans, J. H. Jensen, and J. A. Helpern, Nature Phys. \textbf{7}, 508 (2011).

\bibitem{novikov2} D. S. Novikov, J. H. Jensen, J. A. Helpern, and E. Fieremans, 
Proc. Natl. Acad. Sci. U.S.A. \textbf{111}, 5088 (2014).

\bibitem{grebenkov} D. S. Grebenkov, Rev. Mod. Phys. \textbf{79}, 1077 (2007).

\bibitem{nikol} E. Sykova and C. Nicholson, Physiol. Rev. \textbf{88}, 1277 (2008).

\bibitem{sen-3-2-tail} T. M. de Swiet and P. N. Sen, J. Chem. Phys. \textbf{104}, 206 (1996); 
P. N. Sen, Concepts in Magn. Reson. Part A \textbf{23A}, 1 (2004).

\bibitem{loskut2012} V. V. Loskutov, J. Magn. Reson. \textbf{216}, 192 (2012); 
V. V. Loskutov and V. A. Sevriugin, J. Magn. Reson. \textbf{230}, 1 (2013); 
F. Erdel, M. Baum and K. Rippe, J. Phys.: Condens. Matter \textbf{27}, 064115 (2015).

\bibitem{vincent} V. Tejedor, O. Benichou, R. Voituriez, R. Jungmann, F.
Simmel, C. Selhuber-Unkel, L. Oddershede and R. Metzler, Biophys. J.
\textbf{98}, 1364 (2010); D. O'Malley, V. V. Vesselinov, and J. H. Cushman,
J. Stat. Phys. \textbf{156}, 896 (2014); A. Robson, K. Burrage, and M. C. Leake,
Phil. Trans. R. Soc. B \textbf{368}, 20120029 (2012); H. Kr\"usemann, A. Godec,
and R. Metzler, Phys. Rev. E \textbf{89}, 040101(R) (2014);
S. Condamin, V. Tejedor, R. Voituriez, O.
B{\'e}nichou, and J. Klafter, Proc. Natl. Acad. Sci. \textbf{105}, 5675
(2008).


\end{thebibliography}
\end{document}